\documentclass[prb,twocolumn,showpacs,amsmath,amssymb,floatfix]{revtex4}

\usepackage{graphicx}
\usepackage{dcolumn}
\usepackage{bm}

\begin{document}

\title{\bf High Energy Dynamics of the Single Impurity Anderson Model} 

\author{Carsten Raas$^1$, G\"otz S.~Uhrig$^1$, and Frithjof B.~Anders$^2$}

\affiliation{$^1$Institut f\"ur Theoretische Physik, Universit\"at zu
    K\"oln, Z\"ulpicher Str.\ 77, 50937 K\"oln, Germany}
\affiliation{$^2$Institut f\"ur Theoretische Physik, Universit\"at Bremen,
  28334 Bremen, Germany}
  
\date{\today}

\begin{abstract}
The quantitative control of the dynamic correlations of single impurity 
Anderson 
models is essential in several very active fields. We analyze the one-particle
Green function with a constant energy resolution by dynamic density-matrix
renormalization. In contrast to other approaches, sharp dominant resonances at 
high energies are found. Their origin and importance are discussed.
\end{abstract}


\maketitle

Single impurity models are at the very basis of the description of
strong correlation phenomena. Landmarks are the Kondo problem
\cite{kondo64} and the single impurity Anderson model (SIAM) \cite{ander61},
for a review see Ref.~\onlinecite{hewso93}.

The interest in the quantitative analysis of SIAMs has been intensified
considerably by the advent of a systematic mapping of strongly correlated
lattice models to effective SIAMs supplemented by a self-consistency
condition. This is the key point of dynamic mean-field theory (DMFT)
\cite{jarre92,georg92a} which is based
on an appropriate scaling of the non-local parts of the Hamiltonian
\cite{metzn89a,mulle89a}, for reviews see Ref.~\onlinecite{prusc95,georg96}.
In recent years, the DMFT is applied very successfully in combination
with ab-initio density functional calculations \cite{anisi97,licht98}.
In this way, the unbiased knowledge about the bands could be 
enhanced by the inclusion of interaction effects between the excited
quasi-particles. It turned out that the combination of density
functional results and DMFT makes the quantitative understanding
of spectroscopic data possible \cite{held01b}.

So far, the methods applied to the SIAM
were designed to capture the low-energy physics, in
particular the fixed points and the thermodynamics 
\cite{krish80a,schlo82}. The numerical renormalization group (NRG)
was later extended to calculate also dynamic, i.e., spectral information.
It provides reliable data on the scale of the Kondo temperatures $T_{\rm K}$
, see Ref.~\onlinecite{hewso93,bulla00a} and references therein. 
On larger scales, the energy resolution is less well-controlled.

But in various applications
the behavior at higher energies is important to achieve quantitative
accuracy. For instance, the self-consistency cycle of the 
DMFT mixes modes at all energies. Hence, excellent quantitative control 
over the dynamics at high energies is indispensable, even if finally only 
the behavior at  low energies matters. 

Another application is the  optical control of isolated
 $S=1/2$ impurities or quantum dots coupled to narrow bands.
If the impurities differ so
that the energy between the singly occupied ground state and the
excited double occupancy differs, they can be switched 
selectively  from the ground state to the double occupancy (and back)
by shining light at the resonant frequency  onto the sample.
The life-time of the double occupancy, i.e., the inverse line width of
the resonance,
determines how well the resonance condition has to be met,
how selective the switching can be, and how stable the excited state is.

In view of the above, we perform a numerical investigation
which aims to describe both the low-energy dynamics and the
high-energy dynamics quantitatively. To this end, we use an energy resolution
which is constant for all energies. 
Features at low energies are not as delicately resolved as by NRG,
but in return features at high energies are much better under control.
We apply the dynamic density-matrix renormalization
(D-DMRG) \cite{hallb95,kuhne99a,hovel00}
to compute the one-particle propagator. The DMRG is a real-space
approach \cite{white92,white93,pesch98} which works best for
open boundary conditions so that it is particularly well-suited to 
 treat impurity problems. 

\begin{figure}[htb]
  \begin{center}
    \includegraphics[width=\columnwidth]{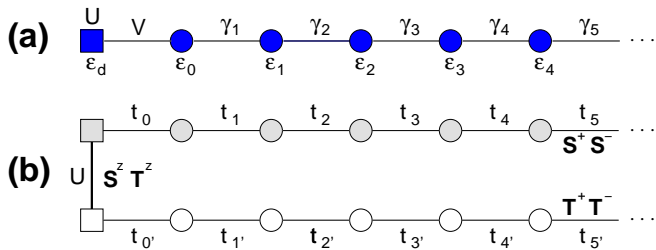}
    \caption{\label{fig:1} 
      (a) Single impurity model with the bath as half-infinite chain. 
      (b) equivalent spin model: $T$-spins come from $\uparrow$-fermions, 
      $S$-spins from $\downarrow$-fermions.}
  \end{center}
\end{figure}

The model investigated at zero temperature is the symmetric Anderson model 
\begin{eqnarray}
\nonumber
&&\mathcal{H} =
\sum_{\sigma}\epsilon_d n_{d,\sigma}
+ U n_{d,\downarrow}n_{d,\uparrow}
+ V \sum_\sigma(d_{\sigma}^{\dagger} c_{0,\sigma}^{\phantom{\dagger}} + 
\text{h.c.})\\
&&
+ \sum_{n,\sigma} \gamma_{n+1}
  \left(c_{n,\sigma}^\dag c_{n+1,\sigma}^{\phantom{\dagger}} + 
\text{h.c.}\right)+ 
\sum_{n,\sigma} 
\epsilon_n c_{n,\sigma}^{\dagger} c_{n,\sigma}^{\phantom{\dagger}} \
\label{hamilton}
\end{eqnarray}
with arbitrary
density of states (DOS) $\rho_0(\omega)$ of the $U=0$ 
 one-particle Green function $G_0(\omega)$ of the
$d$-electron. The parametrization in (\ref{hamilton}) is chosen such
that the coefficients $(\epsilon_n,\gamma_n)$ are the continued
fraction coefficients  of the hybridization function
\begin{equation}
 \label{GAM_confrac}
  {\Gamma}(\omega)=\frac{V^2} {\omega-\epsilon_0-{\displaystyle
  \frac{\gamma_1^2}{\omega-\epsilon_1-{\displaystyle \frac{\gamma_2^2}{\omega
  -\cdots}}}}}\ .
\end{equation}
This model has particle-hole
symmetry iff $\epsilon_d=-U/2$ and $\epsilon_n=0$ for all
$N\ge0$. The representation as continued fraction \cite{viswa94}
(see Fig.~\ref{fig:1}a) is optimum for the DMRG which is designed 
for chains. We look at a generic situation with finite band width
$W=2D$. For simplicity we choose a $\Gamma(\omega)$ with semi-elliptic 
DOS, i.e., $\gamma_n=D/2$. For $V=D/2$, the free DOS $\rho_0(\omega)
=2\sqrt{D^2-\omega^2}/(\pi D^2)$ is also semi-elliptic.

The problem illustrated in Fig.~\ref{fig:1}a is mapped by two standard 
Jordan-Wigner
transformations to two $XY$ spin $1/2$ chains, the $S$-chain and the $T$-chain.
The $S$-chain results from the $\uparrow$ fermions, the $T$-chain
 from the $\downarrow$ fermions. They are coupled at site zero where
the density-density coupling is mapped to the product of $z$-components.
The resulting chain is depicted for the symmetric SIAM in Fig.~\ref{fig:1}b.
The couplings are given by $t_0=t_{0'}=V$ and $t_i=t_{i'}=\gamma_i$
for $i\ge1$. The mapping from fermions to spins
avoids the fermionic Fock space which would imply
numerically difficult long-range effects. The mapping makes
the Hilbert space the direct product of the local Hilbert spaces at 
each site.

The DMRG  can easily determine the
ground state $|0\rangle$ and its energy $E_0$
for a finite chain. So the chain in Fig.~\ref{fig:1}b
is truncated such that there are $L$ spins in the upper and in
the lower part of the chain corresponding
originally to a truncated bath of $L-1$ fermions plus the impurity.
The dynamic quantity we are interested in is the retarded Green
function at zero temperature
\begin{equation}
\label{eq:greendef}
G^>(\omega+i\eta) = \langle 0|
\mathbf{S}_{0}^{-} (\omega+i\eta-(\mathcal{H}-E_0))^{-1} 
\mathbf{S}_{0}^{+} |0\rangle
\end{equation}
where the superscript ${}^>$ 
implies that (\ref{eq:greendef}) represents only the part
 of the usual Green function at non-negative frequencies.
In the symmetric case, the complete function is recovered by
$G(\omega+i\eta)= G^>(\omega+i\eta) - G^>(-\omega-i\eta)$. In the asymmetric
case, $G^<(\omega+i\eta) = \langle 0|
\mathbf{S}_{0}^{+} (\omega+i\eta-(\mathcal{H}-E_0))^{-1} 
\mathbf{S}_{0}^{-} |0\rangle$
must be determined separately whereby $G(\omega+i\eta)= G^>(\omega+i\eta) -
 G^<(-\omega-i\eta)$
is obtained. We stress that $G(\omega+i\eta)$ is the fermionic propagator even
though it is computed in terms of spins after the Jordan-Wigner mapping.

The key idea of the dynamic DMRG is to include the real and the imaginary 
part of the correction vector $|c\rangle$ in the target states of a standard 
DMRG algorithm \cite{kuhne99a,hovel00}.
The natural choice is 
$|c\rangle=(\omega+i\eta-(\mathcal{H}-E_0))^{-1}\mathbf{S}_{0}^{+}
|0\rangle$. The computation of $|c\rangle$ is numerically the most
demanding step due to the inversion of an almost singular non-hermitean matrix.
We prefer to stabilize this inversion by optimized algorithms \cite{freun93a}
instead of using
the variational approach proposed by Jeckelmann \cite{jecke02}
which requires a minimization in a high-dimensional Hilbert space.

The numerical calculations cannot be performed for $\eta=0$. 
Even small values of $\eta$ are very time consuming.
So we  compute first $G(\omega+i\eta)$ at finite $\eta$. The spectral 
density  $\rho^{(\eta)}(\omega) = - \frac{1}{\pi} \Im G(\omega+i\eta)$ can be
seen as the actual spectral density $\rho(\omega)$ convoluted by the 
Lorentzian  $\rho_\text{L}(\omega) =(\eta/\pi)/(\omega^2+\eta^2)$  
of width $\eta$. 
Hence it is possible to retrieve $\rho(\omega)$ by deconvolution. 
A standard technique for deconvolution is Fourier transformation, 
realized best by fast Fourier transforms,
division by $\exp(-\eta\tau)$ plus  low-pass filtering,
and the inverse transform.
A flexible alternative with similar properties is
the explicit matrix inversion of the convolution procedure \cite{gebha03}.

\begin{figure}[htb]
  \begin{center}
    \includegraphics[width=\columnwidth,angle=0]{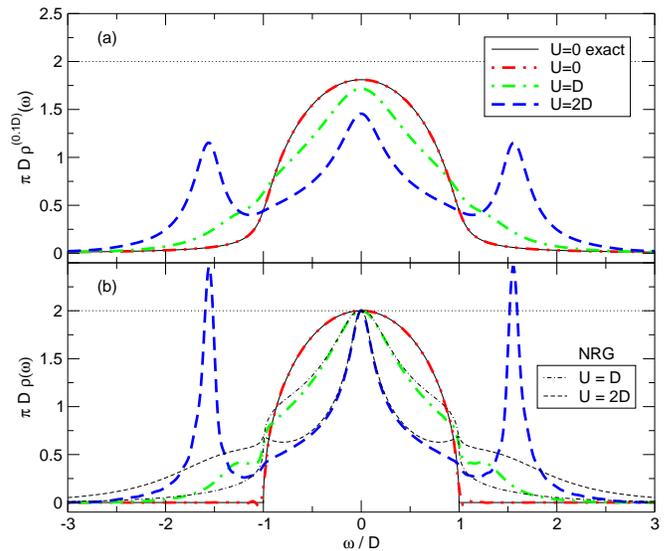}
    \caption{\label{fig:2} (a) Spectral densities for $V=D/2$ broadened by $\eta=0.1D$. 
    Chain length $L=80$ fermionic sites; 128 DMRG states kept. 
      (b) Spectral densities from (a)
      deconvoluted in the time domain. NRG data ($\Lambda=1.8$, 1500 states)
      depicted by thin dashed lines.
      The exact curves represent the analytic $L=\infty$ results.}
  \end{center}
\end{figure}
In Fig.~\ref{fig:2}a, generic broadened spectral densities
are plotted as they are computed by D-DMRG. 
Obviously, the value $\rho^{(\eta)}(0)$ is not
independent of $U$. Increasing the chain length $L$ does not
lead to any significant change in the data (not shown).
Fig.~\ref{fig:2}b displays the deconvoluted data. The deconvolution
works very well except for some slight overshooting in regions where
the spectral density changes rather abruptly. In particular, 
the value $\rho(0)$ is pinned to $D/(2\pi V^2)$ independent of $U$
as required by Friedel's sum rule and the density of states rule
\cite{lutti60b,lutti61,ander91,hewso93}.
We take this fact as convincing evidence for the reliability of the 
numerical algorithm.

The central peak at $\omega=0$ is the Abrikosov-Suhl resonance (ASR).
For larger $U$ (smaller $V$) its width decreases rapidly so that the ASR
is very difficult to resolve \cite{nishi03}
 unless more elaborate deconvolution schemes are devised \cite{raas04a}.
So a quantitative analysis of the ASR is postponed to future work.

For comparison, the thin dashed lines in Fig.~\ref{fig:2}b depict standard NRG
data \cite{bulla98,bulla00a}. For small frequencies NRG is well-controlled. 
Indeed, for $|\omega|\lessapprox D/3$, NRG and D-DMRG data agree excellently
lending further support to the D-DMRG approach. 
Outside the band,  the NRG spectra appear to be too wide due to the
chosen constant broadening on a logarithmic mesh. This
broadening does not account for the absence of states outside the bare band.
The NRG does not possess intrinsic information about the peak widths.
The position of the high energy peak in
the raw NRG data, however, coincides with the D-DMRG result.

An increase in $U$ leads to the formation
of Hubbard satellites below and above the free band 
(Fig.~\ref{fig:2}). They are situated at energies 
$\omega_{\rm up/low} = \pm(U/2+ \delta_{\rm shift}),\delta_{\rm shift}>0 $
and become more pronounced on increasing $U$ in two ways. They capture
more weight and they become sharper. For $T_\text{K}\to 0$
the weight reaches 1/2, see  Ref.\ \onlinecite{ander91}. 
The sharpening has not 
been discussed quantitatively before although the extended non-crossing
approximation \cite{prusc89} provides sharp satellites if
they  lie outside the bare bands, see e.g.\ Fig.~1 in Ref.~\onlinecite{ander91}.
Recently, indications have 
occured \cite{gebha03} that other standard algorithms
overestimate the width of the Hubbard satellites.
The exaggerated width of the NRG data at high energies
results from the Gaussian broadening of the order of the
energy range \cite{sakai89}.

\begin{figure}[htb]
  \begin{center}
    \includegraphics[width=\columnwidth,angle=0]{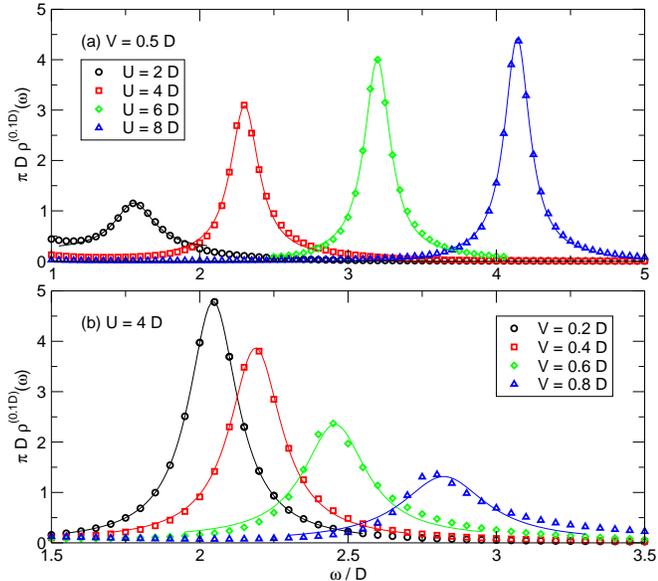}
    \caption{\label{fig:3} D-DMRG data of the upper Hubbard
      satellite at $\eta=0.1D$. Thin lines are fits with
      Lorentzians and an offset
      $\propto \eta_{\rm eff}/(\eta_{\rm eff}^2 +(\omega-\omega_{\rm up})^2)+C$.
      The fits were done for the intervals shown.
      (a) dependence on $U$ at constant $V$ ($L=80$); 
      (b) dependence on $V$ at constant $U$ ($L=40$).
      }
  \end{center}
\end{figure}

To investigate the line shapes of the satellites
we plot them for various values of $U$ and $V$
in Fig.~\ref{fig:3}. The ASR at $\omega=0$
is not displayed since it is too much smeared out at $\eta=0.1D$
for larger values of the interaction.
The shifts $\delta_{\rm shift}$ increase on increasing $V$;
they decrease on growing interaction $U$. The widths behave
qualitatively similar. A complete deconvolution suffers
unfortunately from severe overshooting due to the sharpness
of the resonance.
To make the analysis nonetheless quantitative we fit
the broadened data by Lorentzians  plus an offset (Fig.~\ref{fig:3}).
These fits work very well for large values of $U$ and  not
too large values of $V$. 

\begin{figure}[htb]
  \begin{center}
    \includegraphics[width=\columnwidth,angle=0]{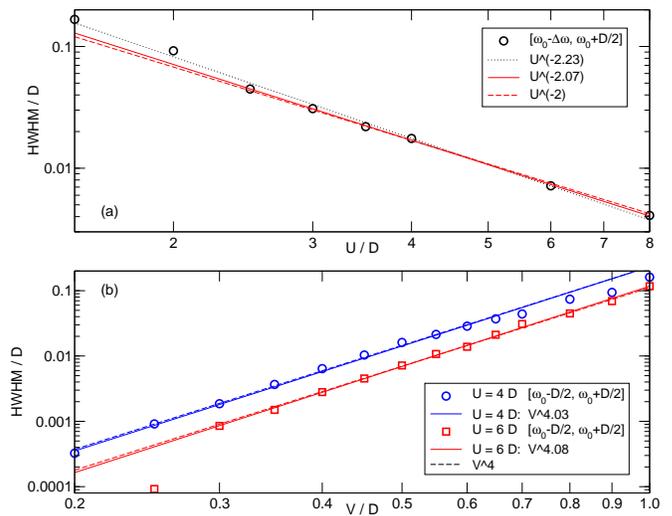}
    \caption{\label{fig:4} Widths (symbols) of the Hubbard satellites
    as found from the fits in Fig.~\ref{fig:3}. The fit intervals
      are given in the legend. The lines show various power law fits.
      (a) dependence on $U$ at $V=D/2$; (b) dependence on $V$.}
  \end{center}
\end{figure}
To deduce the true width of the Hubbard satellite we assume that
it is well described by a Lorentzian. The width $\eta_{\rm eff}$
 of the convolution of two Lorentzians of widths $\eta_1$ and $\eta_2$
is $\eta_{\rm eff} =\eta_1 +\eta_2$. 
From the effective widths $\eta_{\rm eff}$
we deduce the true half-widths at half-maximum (HWHM) of the Hubbard
satellite  by subtracting the artificial broadening $\eta$, i.e.,
HWHM $=\eta_{\rm eff}-\eta$. 
In Fig.~\ref{fig:4}, the widths are depicted as function of
$U$ and of $V$. The results show that the
HWHM are proportional to $V^4/U^2$.
The deviations for smaller widths must be attributed to the numerical
constraints, e.g., finite $\eta$ and finite chain length $L$. 
The deviations for larger widths, mainly for larger values of $V$ and
smaller values of $U$ result from the vicinity of the bare bands.
\begin{figure}[htb]
  \begin{center}
    \includegraphics[width=\columnwidth,angle=0]{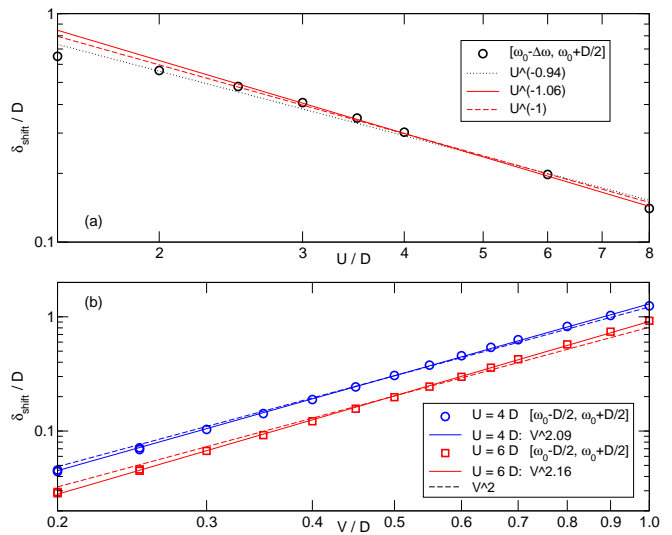}
    \caption{\label{fig:5} Position shifts of the Hubbard satellites
    analyzed like the widths in Fig.~\ref{fig:4}; (a) $V=D/2$; (b) see legend.}
  \end{center}
\end{figure}
Fig.~\ref{fig:5} displays the analogous analysis for the shifts $\delta_{\rm shift}$
of the satellite positions. Again, strong evidence for
power law behavior is found, namely $\delta_{\rm shift}\propto V^2/U$.

How can the above findings be understood? Let us start by 
the positions. The energy levels of isolated impurities, i.e.,
$V=0$ are at $\pm U/2$ \cite{schri66}. 
Switching on $V$ mixes the impurity levels with
the bath states which lie in the interval $[-D,D]$. If $U$ is 
large compared to $D$
second order perturbation in $V$ implies that the impurity levels are
repelled from the bath states. The shift should be of the order
of $J=8V^2/U$, see Eq.~(11) in Ref.~\onlinecite{schri66},
which agrees nicely with the power laws in Fig.~\ref{fig:5}.

The widths of the satellites have been discussed quantitatively
when they lie within the bare band \cite{logan98}. If the
satellites lie outside, perturbation theory
in $U$, to second order or the random phase approximation,
 implies that a finite width is to be expected at least for 
$U<6D$. But the reasoning in powers of $U$ is not helpful for
$U>6D$ and it does not explain the power laws found. So we return
to perturbing in powers of $V$. The impurity levels mix
with particle-hole excitations in the bands, see Eq.~(10a) in
Ref.~\onlinecite{schri66}.  
In the {\em symmetric} case the doubly occupied electron
and hole state are degenerate so that mixing with 
particle-particle (or hole-hole) states matters also, see Eq.~(12) in
Ref.~\onlinecite{schri66}. The mixing is of order $J=8V^2/U$.
So Fermi's golden rule implies a 
life time of $J^2 N_0$ where $N_0$ measures the density of
states with which the impurity level mixes; $N_0$ is of the order of $D^{-1}$.
Indeed, HWHM $\propto J^2$ explains conclusively the 
data of Fig.~\ref{fig:4}.

So far, the width of the Hubbard satellites for $U>2D$ 
was extracted under the
assumption that the satellites are Lorentzians. 
Further investigations of the line shape
are urgently called for. Numerically, improvements of the resolution
are necessary to determine the line shape of the 
satellites explicitly. Analytically, the quantitative argument 
for the widths must be supplemented by
an explicit calculation of the line shapes for $V^2/U\to 0$.

In summary, we have investigated the dynamic propagator 
of the SIAM by D-DMRG. This powerful large-scale algorithm 
provides information with a constant energy resolution. 
Up to moderate interactions $U\approx 2D$, 
deconvolution yields the explicit spectral densities.
For larger interactions, the width of sharp resonances can
be extracted by fitting Lorentzians.
In particular, we analyzed the positions and widths of the Hubbard
satellites. The shifts are of order $V^2/U$ due to level repulsion;
the line widths are of order $V^4/U^2$. 

Especially the sharpness of Hubbard peaks is missed by other
zero temperature algorithms for the SIAM. 
Hence the D-DMRG is a very valuable complementary tool.
Position and width of the Hubbard satellites are important for
several widely used applications, e.g., in the self-consistency
cycle of the DMFT.

{\it Acknowledgements} \quad
We thank R.~Bulla, L.~Craco, M.~Gr\"uninger,
 T.~H\"ovelborn, D.~Logan, H.~Monien,
 and E.~M\"uller-Hartmann for helpful discussions
 and the DFG  for financial support (Uh 90/3-1 and SFB 608).


\begin{thebibliography}{10}

\bibitem{kondo64}
J. Kondo, Prog. Theor. Phys. {\bf 32},  37  (1964).

\bibitem{ander61}
P.~W. Anderson, Phys. Rev. {\bf 124},  41  (1961).

\bibitem{hewso93}
A.~C. Hewson, {\em The Kondo Problem to Heavy Fermions} (Cambridge University
  Press, Cambridge, 1993).

\bibitem{jarre92}
M. Jarrell, Phys. Rev. Lett. {\bf 69},  168  (1992).

\bibitem{georg92a}
A. Georges and G. Kotliar, Phys. Rev. B {\bf 45},  6479  (1992).

\bibitem{metzn89a}
W. Metzner and D. Vollhardt, Phys. Rev. Lett. {\bf 62},  324  (1989).

\bibitem{mulle89a}
E. M\"uller-Hartmann, Z. Phys. B {\bf 74},  507  (1989).

\bibitem{prusc95}
T. Pruschke, M. Jarrell, and J.~K. Freericks, Adv. Phys. {\bf 44},  187
  (1995).

\bibitem{georg96}
A. Georges, G. Kotliar, W. Krauth, and M.~J. Rozenberg, Rev. Mod. Phys. {\bf
  68},  13  (1996).

\bibitem{anisi97}
V.~I. Anisimov {\it et~al.}, J. Phys.: Condens. Matter {\bf 9},  7359  (1997).

\bibitem{licht98}
A.~I. Lichtenstein and M. I. Katsnelson, Phys. Rev. B {\bf 57},  6884  (1998).

\bibitem{held01b}
K. Held {\it et~al.}, Int. J. Mod. Phys. B {\bf 15},  2611  (2001).

\bibitem{krish80a}
H.~R. Krishna-murthy, J.~W. Wilkins, and K.~G. Wilson, Phys. Rev. B {\bf 21},
  1003 and 1044 (1980).


\bibitem{schlo82}
P. Schlottmann, Z. Phys. B {\bf 49},  109  (1982).

\bibitem{bulla00a}
R. Bulla,  in {\em Advances in Solid State Physics}, edited by B. Kramer
  (Vieweg Verlag, Braunschweig, 2000)  {\bf 40}, 129.

\bibitem{hallb95}
K.~A. Hallberg, Phys. Rev. B {\bf 52},  9827  (1995).

\bibitem{kuhne99a}
T.~D. K\"uhner and S.~R. White, Phys. Rev. B {\bf 60},  335  (1999).

\bibitem{hovel00}
T. H\"ovelborn, 
diploma thesis, Bonn/K\"oln, 2000;
available at www.thp.uni-koeln.de/$\widetilde{\phantom{w}}${}gu.

\bibitem{white92}
S.~R. White, Phys. Rev. Lett. {\bf 69},  2863  (1992).

\bibitem{white93}
S.~R. White, Phys. Rev. B {\bf 48},  10345  (1993).

\bibitem{pesch98}
I. Peschel, X. Wang, M. Kaulke, and K. Hallberg, {\em Density-Matrix
  Renormalization}, Vol.~528 of {\em Lecture Notes in Physics} (Springer,
  Berlin, 1999).

\bibitem{viswa94}
V.~S. Viswanath and G. M\"uller, {\em The Recursion Method; Application to
  Many-Body Dynamics}, Vol.~m23 of {\em Lecture Notes in Physics}
  (Springer-Verlag, Berlin, 1994).

\bibitem{freun93a}
R.~W. Freund, SIAM J. Sci. Comput. {\bf 14}, 470 (1993).

\bibitem{jecke02}
E. Jeckelmann, Phys. Rev. B {\bf 66},  045114  (2002).

\bibitem{gebha03}
F. Gebhard {\it et~al.}, Eur. Phys. J. B in press, cond-mat/0306438.

\bibitem{lutti60b}
J.~M. Luttinger, Phys. Rev. {\bf 119},  1153  (1960).

\bibitem{lutti61}
J.~M. Luttinger, Phys. Rev. {\bf 121},  942  (1961).

\bibitem{ander91}
F.~B. Anders, N. Grewe, and A. Lorek, Z. Phys. B {\bf 83},  75  (1991).

\bibitem{nishi03}
S. Nishimoto and E. Jeckelmann, J. Phys.: Cond. Matter in press, 
cond-mat/0311291.

\bibitem{raas04a}
C. Raas and G.S. Uhrig, in preparation

\bibitem{bulla98}
R. Bulla, A.~C. Hewson, and T. Pruschke, J. Phys.: Condens. Matter {\bf 10},
  8365  (1998).

\bibitem{prusc89}
T. Pruschke and N. Grewe, Z. Phys. B {\bf 74},  439  (1989).

\bibitem{sakai89}
O. Sakai, Y. Shimizu, and T. Kasuya, J. Phys. Soc. Jpn. {\bf 58},  3666  (89).

\bibitem{schri66}
J.~R. Schrieffer and P.~A. Wolff, Phys. Rev. {\bf 149},  491  (1966).

\bibitem{logan98}
D.~E. Logan, M.~P. Eastwood, and M.~A. Tusch, J. Phys.: Condens. Matter {\bf
  10},  2673  (1998).

\end{thebibliography}

\end{document}